\documentclass[conference]{IEEEtran}

\usepackage{graphicx}
\usepackage{tabularx}

\usepackage[T1]{fontenc}
\usepackage[utf8]{inputenc}

\usepackage[textsize=scriptsize]{todonotes}
\usepackage{csquotes}
\usepackage{nicefrac}
\usepackage{tikz}
\usetikzlibrary{patterns}
\usetikzlibrary{shapes.geometric, positioning, quantikz}
\usepackage{hyperref}
\usepackage{amssymb}
\usepackage{amsthm}
\usepackage{booktabs}
\usepackage{flushend}

\usepackage[font=footnotesize]{subcaption}
\usepackage{yquant}
\useyquantlanguage{groups}
\usepackage{pgfplots}
\usepackage{xurl}

\newtheorem{example}{Example}

\usepackage[style=ieee, maxnames=6, minnames=1, date=year, doi=false,isbn=false,backend=biber]{biblatex}
\AtEveryBibitem{
  \clearname{editor}%
  \clearfield{series}%
  \clearfield{isbn}%
  \clearfield{issn}%
  \clearfield{volume}%
  \clearfield{number}%
  \clearfield{pages}%
  \clearfield{doi}%
  \clearfield{urldate}%
}

\DeclareFieldFormat
  [misc, book]
  {title}{\mkbibquote{#1}}

\begin{document}
\title{Utilizing Resource Estimation for the \\Development of Quantum Computing Applications}

\author{
	\IEEEauthorblockN{Nils Quetschlich\IEEEauthorrefmark{1}\hspace*{1.5cm}Mathias Soeken\IEEEauthorrefmark{2}\hspace*{1.5cm}Prakash Murali\IEEEauthorrefmark{3}\hspace*{1.5cm}Robert Wille\IEEEauthorrefmark{1}\IEEEauthorrefmark{4}}
	\IEEEauthorblockA{\IEEEauthorrefmark{1}Chair for Design Automation, Technical University of Munich, Germany}
	\IEEEauthorblockA{\IEEEauthorrefmark{2}Microsoft Quantum, Switzerland}
	\IEEEauthorblockA{\IEEEauthorrefmark{3}University of Cambridge, United Kingdom}
	\IEEEauthorblockA{\IEEEauthorrefmark{4}Software Competence Center Hagenberg GmbH (SCCH), Austria}
	\IEEEauthorblockA{\href{mailto:nils.quetschlich@tum.de}{nils.quetschlich@tum.de}\hspace{1cm}\href{mailto:mathias.soeken@microsoft.com}{mathias.soeken@microsoft.com}\hspace{1cm}\href{mailto:pm830@cam.ac.uk}{pm830@cam.ac.uk}\hspace{1cm} \href{mailto:robert.wille@tum.de}{robert.wille@tum.de}\\
	\url{https://www.cda.cit.tum.de/research/quantum} \hspace{1cm} \url{https://quantum.microsoft.com}
	}
}

\maketitle

\begin{abstract}
Quantum computing has made considerable progress in recent years in both software and hardware. 
But to unlock the power of quantum computers in solving problems that cannot be efficiently solved classically, quantum computing at scale is necessary. Unfortunately, quantum simulators suffer from their exponential complexity and, at the same time, the currently available quantum computing hardware is still rather limited (even if roadmaps make intriguing promises).
Hence, in order to evaluate quantum computing applications, end-users are still frequently restricted to toy-size problem instances (which additionally often do not take error correction into account). This substantially hinders the development and assessment of \mbox{real-world} quantum computing applications. In this work, we demonstrate how to utilize \emph{Resource Estimation} to improve this situation. We show how the current workflow (relying on simulation and/or execution) can be complemented with an estimation step, allowing that \mbox{end-users}
(1)~actually can consider \mbox{real-world} problem instances already today (also considering error correction schemes and correspondingly required hardware resources),
(2)~can start exploring possible optimizations of those instances across the entire design space, and
(3)~can incorporate hypotheses of hardware development trends to derive more informed and, thus, better design space parameters.
Overall, this enables \mbox{end-users} already today to check out the promises of possible future quantum computing applications, even if the corresponding hardware to execute them is not available yet.

\end{abstract}

\begin{figure*}[t]
\centering
\resizebox{0.90\linewidth}{!}{
				\begin{tikzpicture}
				
				\node (p) at (-2,0) {
				
\resizebox{0.12\linewidth}{!}{
				\begin{tikzpicture}
	 \node[shape=circle,draw=black, line width=1mm] (1) at (0,0) {H};
    \node[shape=circle,draw=black, line width=1mm] (2) at (1,0) {H};

    \path [color=black, draw, line width=1mm] (1) edge node[] {}(2);
				\end{tikzpicture}
				}};
				
				\node [right=of p, align=center](designspace)  {
				\begin{tikzpicture}
	 \node(a){Mathmatical \\Formulation};
	 \node[below=of a, yshift=5mm](b) {Algorithm \\Selection};
	 \node[below=of b, yshift=5mm](c) {Encoding};
	\draw [-latex, line width=0.5mm] (a) -- (b);
	\draw [-latex, line width=0.5mm] (b) -- (c);
	\node[fit=(a)(c), draw] {};
				\end{tikzpicture}
				};
				
				\node [right=of designspace](qc)  {
\resizebox{0.22\linewidth}{!}{
				\begin{tikzpicture}
				\begin{yquant}

						qubit {$q_1$} q;
						qubit {$q_2$} q[+1];
						qubit {$q_3$} q[+1];
						qubit {$q_4$} q[+1];

                        box {$R_Y$} q[0-3];
                        cnot q[1] | q[0];
                        cnot q[2] | q[1];
                        cnot q[3] | q[2];
                        box {$R_Y$} q[0-3];
				  \end{yquant}

				\end{tikzpicture}}};
				
				\node [right=of qc](dev){

\resizebox{0.15\linewidth}{!}{
			
				\begin{tikzpicture}    
				\node[shape=circle,draw=black] (Q1) at (0,0) {$Q_1$};
    \node[shape=circle,draw=black] (Q2) at (1,0) {$Q_2$};
    \node[shape=circle,draw=black] (Q3) at (2,0) {$Q_3$};
    \node[shape=circle,draw=black] (Q4) at (1,-1) {$Q_4$};
    \node[shape=circle,draw=black] (Q5) at (1,-2) {$Q_5$};

	\draw (Q1) edge (Q2);
	\draw (Q2) edge (Q3);
	\draw (Q2) edge (Q4);
	\draw (Q4) edge (Q5);
				
				\end{tikzpicture}}};

				\node [right=of dev](res){
				\resizebox{0.16\linewidth}{!}{
				\begin{tikzpicture}
\begin{axis}[
    title=\Huge{$H_2$ Energy},
    xmin=0, xmax=8.5,
    ymin=0, ymax=0.9,
    ticks=none,
]

\addplot[
    color=blue,
    line width=1.5mm,
    ]
    coordinates {
    (1,0.8)(1.5,0.2)(2,0.1)(2.5,0.2)(3,0.25)(4,0.28)(5,0.3)(6, 0.33)(7,0.34)(8,0.35)
    };
    
\end{axis}
\end{tikzpicture}}};
				
	\node[fit=(dev)(res), draw, label=Execution](exec) {};
				
    \draw[-latex, line width=0.5mm] (p.east)  -- (designspace.west);
    \draw[-latex, line width=0.5mm] (designspace.east)  -- (qc.west);
    \draw[-latex, line width=0.5mm] (qc.east)  -- (exec.west);
    \draw[-latex, line width=0.5mm] (dev.east)  -- (res.west);
    
	\node[fit=(dev)(res), draw, label=Execution](exec) {};
				\end{tikzpicture}

}
	
	\vspace*{-5mm}
	\hspace{-4mm}\subfloat[Problem.\label{fig:wf1_1} ]{\hspace{.18\linewidth}}
	\hspace{-5mm}\subfloat[Design Space.\label{fig:wf1_2}]{\hspace{.18\linewidth}}
	\hspace{3mm}
	\subfloat[Quantum Program.\label{fig:wf1_3} ]{\hspace{.18\linewidth}}
	\hspace{3mm}
	\subfloat[Device Selection.\label{fig:wf1_4} ]{\hspace{.18\linewidth}}
	\hspace{0mm}
	\subfloat[Result.\label{fig:wf1_5} ]{\hspace{.16\linewidth}}
	
	\caption{Current workflow for the development of quantum computing applications.}
	\label{fig:workflow}
 \vspace{-6mm}
\end{figure*}
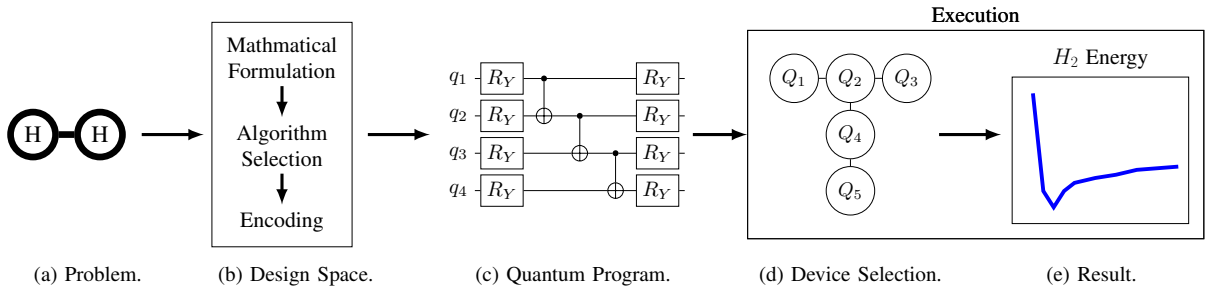  

\section{Introduction}\label{sec:intro}

In recent years, quantum computing has witnessed remarkable advancements in both software and hardware. As a result, an expanding range of quantum devices with progressively enhanced qubit quality has become available, leading to increased interest in academia and industry for tackling diverse problems across multiple application domains. This trend is indicative of the growing potential of quantum computing to offer transformative solutions to complex problems that are beyond the capabilities of classical computing~\cite{scholten2024assessing}.

The utilization of quantum computers to solve complex problems involves a \mbox{multi-stage} process:
The first step involves selecting or developing a quantum algorithm that offers a quantum advantage, i.e., can solve the problem \emph{better} compared to the \mbox{best-known} classical algorithm (e.g., in terms of algorithmic complexity, execution time, or solution quality).
The second step entails the encoding of the problem in terms of a quantum program that can be compiled into machine instructions for a quantum computer. 
This process factors in various constraints on quantum gate sets and qubit connectivities that arise from different quantum computing technologies being explored, such as superconducting~\cite{kjaergaard2020superconducting}, ion traps~\cite{bruzewicz2019trapped}, neutral atoms~\cite{henriet2020quantum}, or Majorana qubits~\cite{majorana}. 
The third step involves executing the compiled program on the quantum computer (or a corresponding simulator), and the final step covers the interpretation of the measurement results from the quantum computer as the solution to the original problem.
Developing such solutions is an active area of research in domains such as  finance~\cite{stamatopoulosOptionPricingUsing2020}, chemistry~\cite{peruzzoVariationalEigenvalueSolver2014}, machine learning~\cite{zoufalQuantumGenerativeAdversarial2019}, and optimization~\cite{harwoodFormulatingSolvingRouting2021}.

Unfortunately, this process fails as soon as practically relevant applications are considered. 
This is because the application of the described workflow currently relies on the use of quantum simulators and \mbox{near-term} quantum computers. 
The former deploy classical computers to simulate quantum algorithms---a computationally complex task which is limited to a few dozen qubits. 
The latter supports a larger number of qubits, but their scalability is limited as well. 
Importantly, the noise rates in these \mbox{near-term} quantum computers currently are still too high to support \emph{high-fidelity} executions of practically relevant applications. 

Consequently, a comprehensive evaluation of the potential of quantum computing for a considered problem necessitates the consideration of quantum error correction to scale to larger problem instances closer to real-world problem sizes.
This induces an overhead in the required resources (such as, e.g., the number of qubits) which effectively exceeds the limits of current devices by far.
All of that could lead to a situation in which certain quantum algorithms might not provide any practical quantum advantage once the presumed characteristics of the \mbox{fault-tolerant} quantum computer and the efficacy of error correction protocols are factored in~\cite{hoefler2023disentangling}. 
Because of this, it is often still not clear what applications are suitable to be solved using quantum computing---constituting a major bottleneck in the progress of quantum computing application development.

Recently, a complementary approach to the execution on simulators or quantum computing hardware emerged: \emph{Resource Estimation}~(RE) such as, e.g., proposed in~\cite{beverland2022assessing,hoefler2023disentangling}.
Instead of actually executing a given quantum program, RE gives an estimate of the resources necessary to execute it in a \mbox{fault-tolerant} fashion.
Although this approach does not return the actual solution, it allows one to determine an early resource estimate---a procedure that has been exploited in conventional computing for decades, e.g., in classical \mbox{HW/SW} \mbox{co-design} where cost estimates are used to guide which functionalities are realized in hardware and which in software (see, e.g.,~\cite{flash2020choi, ieeetc2020_1}).

This paper illustrates how the current workflow to solve problems using quantum simulators and noisy \mbox{near-term} quantum computers can be complemented by substituting the actual \mbox{execution} step with an \mbox{estimation} step---leading to the following three improvements:
Firstly, \mbox{end-users} are enabled to consider \mbox{real-world} problem instances already today (also considering error correction schemes to explore the hardware requirements of a chosen set of design space parameters and determine the required hardware resources). 
These estimates can then be compared with the quantum hardware vendor's roadmaps to obtain a sense of the chosen approach's feasibility.
Secondly, \mbox{end-users} can start exploring optimizations across the entire design space of a quantum computing application and, by that, develop and facilitate \mbox{trade-offs} between different design space parameters.
Thirdly, \mbox{end-users} can even incorporate various hypotheses of hardware development trends and their potential improvements into the methodology. 

On the basis of that, more informed and, thus, better design space parameters are derived.
Overall, the workflow complementation eventually provides valuable insights for the development of future quantum technologies and promising applications for them due to its versatility and applicability across different stages of quantum computing application development.
In other words, this enables \mbox{end-users} already today to check out the promises of possible future quantum computing applications, even if the corresponding hardware to execute them is not available yet.

The remainder of this work is structured as follows:
\autoref{sec:motivation} describes the current workflow to solve problems using quantum computing.
\autoref{sec:proposed} outlines how resource estimation can complement the flow in the absence of \mbox{large-scale} quantum computers.
Then, the complemented workflow is applied in a representative case study in \autoref{sec:use_cases} based on a \mbox{real-world} problem instance from the domain of chemistry to demonstrate its advantages and is discussed in \autoref{sec:discussion}.
\autoref{sec:conclusion} concludes this work.

\section{Motivation}\label{sec:motivation}
In this section, the quantum solution workflow is reviewed and illustrated with a running example.

\subsection{Quantum Solution Workflows}\label{sec:current_wf}
The workflow to solve a problem using quantum computing is illustrated in \autoref{fig:workflow} (based on the workflow proposed in \cite{quetschlich2023mqtproblemsolver}).
Starting with the problem itself as sketched in \autoref{fig:wf1_1}, it must be translated into a form suitable for quantum computing as indicated in \autoref{fig:wf1_2}.
This comprises (1) the mathematical problem formulation, (2) the selection of a quantum algorithm that is generally capable of solving the problem considered, and (3) its encoding into a quantum program based on that---requiring multiple design decisions that eventually form the \emph{design space}.
Any combination of suitable design choices leads to a quantum program as shown in \autoref{fig:wf1_3}.

\begin{example}\label{ex:problem}
A prominent example application in the domain of chemistry is to calculate the \emph{ground state energy} of a molecule such as the $H_2$ molecule illustrated in \autoref{fig:wf1_1}.
To mathematically describe the problem, its \emph{Hamiltonian} must be derived.
This is usually done with the help of already existing software tools, such as \emph{PySCF}~\cite{sun2017pythonbased}.
Subsequently, the resulting Hamiltonian must be mapped to a format suitable for quantum computing, e.g., using the \emph{Jordan-Wigner} mapping. 
Next, the \emph{Variational Quantum Eigensolver}~(VQE,~\cite{peruzzoVariationalEigenvalueSolver2014}) is selected as the solving algorithm since it is one of the prime algorithms for this kind of problem.
For VQE, a respective \emph{ansatz} must be chosen for the encoding as a quantum program---in this case, the \emph{TwoLocal} ansatz\footnote{See \url{https://qiskit.org/documentation/stubs/qiskit.circuit.library.TwoLocal.html} for details.} with a linear entanglement pattern is chosen, resulting in the quantum program being sketched in \autoref{fig:wf1_3}.
\end{example}

Then, the resulting quantum program must be executed.
To this end, both quantum simulators (based on different data structures~\cite{dac2022_special_session} such as, e.g., decision diagrams~\cite{zulehnerAdvancedSimulationQuantum2019, 10.1145/3514355, viamontesImprovingGatelevelSimulation2003, noiseaware_grurl_2022, iccad2019_2}, tensor networks~\cite{biamonte2017tensor, markovSimulatingQuantumComputation2008}, \mbox{matrix-product} states~\cite{ORUS2014117, garciaMatrixProductState2007}, or sparsity~\cite{jaques2021leveragingstatesparsityefficient})
and \mbox{near-term} quantum computers (based on various technologies such as, e.g., superconducting~\cite{kjaergaard2020superconducting}, ion traps~\cite{bruzewicz2019trapped}, neutral atoms~\cite{henriet2020quantum}, or Majorana qubits~\cite{majorana}) are available and a representative must be selected as shown in \autoref{fig:wf1_4}.
This representative must be capable of executing the quantum program---both in terms of the number of required qubits and also on its qubits' characteristics such as the gate execution error rates or decoherence time.

\begin{example}\label{ex:exec}
The quantum program derived in \autoref{ex:problem} requires four qubits.
Therefore, the \emph{ibmq\_perth} device with $5$ qubits is selected which has a limited connectivity as depicted in \autoref{fig:wf1_4}.
\end{example}

To eventually determine the desired solution to the initial problem, the encoded quantum program can now be executed on the chosen device.
Afterwards, the result is extracted from the measurement results.

\begin{example}
When executing the program resulting from \autoref{ex:problem} on the device chosen from \autoref{ex:exec}, the ground state energy is approximated as insinuated in~\autoref{fig:wf1_5}. 
Determining the minimum value of this graph returns the desired value of the $H_2$ molecule.
\end{example}

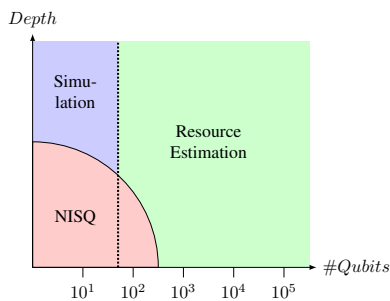
\begin{figure}[t]
\centering
\resizebox{0.60\linewidth}{!}{
\begin{tikzpicture}
\def\height{4.5}

\fill [blue!20!white] (0.0,0.0) rectangle (1.7,\height);
\fill [green!20!white] (1.7,0.0) rectangle (5.5,\height);
\filldraw[fill=red!20!white] (2.5,0) arc[start angle=0, end angle=90, radius=2.5, red] -- (0,2.5) -- (0,0) -- cycle;
\draw [densely dotted, line width=0.4mm] (1.7,  0) -- (1.7,\height) node  {};

\draw [->, -latex] (0,  0) -- (5.65,0) node [right]  {$\#Qubits$};
\draw [->, -latex] (0,0) -- (0,\height+0.15) node [above]  {$Depth$};

\draw [] (1.0,0) -- (1.0,-0.2) node [below]  {$10^1$};
\draw [] (2.0,0) -- (2.0,-0.2) node [below]  {$10^2$};
\draw [] (3.0,0) -- (3.0,-0.2) node [below]  {$10^3$};
\draw [] (4.0,0) -- (4.0,-0.2) node [below]  {$10^4$};
\draw [] (5.0,0) -- (5.0,-0.2) node [below]  {$10^5$};

\node [align=center] at (3.5,2.5){Resource\\ Estimation};
\node [align=center] at (0.85,3.5){Simu-\\lation};
\node [align=center] at (0.85,1){NISQ};

\end{tikzpicture}}
\caption{Current execution limits for quantum programs.}
\label{fig:limits}
\vspace{-5mm}
\end{figure}

\subsection{Current Limitations and General Idea}\label{sec:limits}
There are two options to execute quantum programs: Either the currently available quantum simulators or \mbox{near-term} quantum computers.
However, both options are limited in their capabilities as illustrated in \autoref{fig:limits} and do not provide means for exploring practically relevant problems.

Quantum simulators provide ideal qubits but have a rather small capacity---usually limited to a few dozen qubits---while, on the other hand, the currently available \mbox{so-called} \emph{Noisy Intermediate Scale Quantum}~(NISQ) computers~\cite{Preskill_2018} with up to several hundreds of qubits do not provide a sufficient qubit quality for a reliable execution (and, thus, effectively restricting the quantum program depth).
As a consequence, the development of quantum computing applications at the moment considers mostly \mbox{toy-size} problem instances whose quantum computing solutions still fit the current simulators and NISQ computers---not considering the scalability to larger problem sizes closer to \mbox{real-world} scenarios.

Most quantum device vendors have published roadmaps to significantly scale their device capabilities over the next years.
However, it would be disadvantageous to wait for the availability of sufficiently large quantum devices before considering \mbox{real-world} problem sizes when developing quantum computing applications---leading to a situation where, in the worst case, powerful devices are available, but no suitable applications are available to make use of them.

\subsection{Resource Estimation}\label{sec:re}
\emph{Resource Estimation}~(RE) is a promising approach to overcome this bottleneck and to provide guidance on implementing important quantum algorithm instances today for the hardware of tomorrow.
Instead of executing a given quantum program, RE estimates the required resources (such as the number of qubits and the runtime) based on assumed hardware characteristics such as gate execution times, fidelity rates, and, the underlying error correction scheme.
This methodology allows one to consider quantum programs orders of magnitude larger than the current limits of both quantum simulators and computers.
Although this obviously does not result in an actual execution and, hence, solution, it already gives insights into the applicability and scalability of the chosen design space parameters and assumed hardware characteristics.

RE is an emerging topic within the quantum computing community and, recently, various methodologies have been proposed.
While some works (such as, e.g.,~\cite{litinski2023compute}) focus on rather manual generation of resource estimates, more and more software tools are proposed to automate this procedure---e.g., Microsoft's Azure Quantum Resource Estimator~\cite{beverland2022assessing, vandam2023using}, Google's Qualtran~\cite{QLRnQualtranDocumentation}, Zapata's BenchQ~\cite{Benchq2023}, and MIT Lincoln Lab's pyLIQTR~\cite{PyLIQTR2023}. 
These software tools provide simple means of estimating what resources are necessary to reliably execute a given quantum program based on assumed qubit hardware characteristics.

\section{Resource Estimation-driven Development}\label{sec:proposed}
In this work, we describe how resource estimation complements the quantum solution workflow to allow the development of quantum computing applications for practically relevant problems already today, even when fault-tolerant quantum hardware is not available yet.

\begin{figure*}[t]
\centering
\resizebox{0.9\linewidth}{!}{
				\begin{tikzpicture}
				\node (p) at (-2,0) {
                \includegraphics[width=0.2\linewidth]{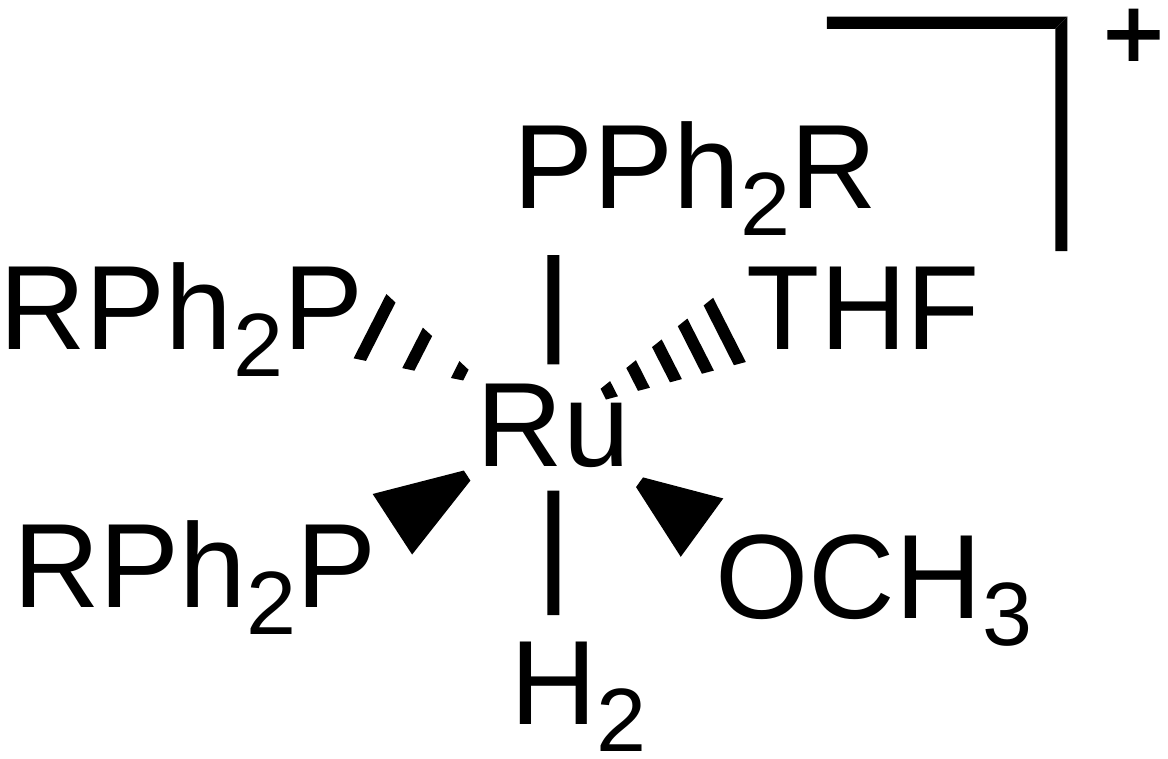}
    };
				\node [right=of p, align=center](designspace)  {

				\begin{tikzpicture}
	 \node(a){Mathmatical \\Formulation};
	 \node[below=of a, yshift=5mm](b) {Algorithm \\Selection};
	 \node[below=of b, yshift=5mm](c) {Encoding};
	\draw [-latex, line width=0.5mm] (a) -- (b);
	\draw [-latex, line width=0.5mm] (b) -- (c);
	\node[fit=(a)(c), draw] {};
				\end{tikzpicture}
				};
				
				\node [right=of designspace](qc)  {
\resizebox{0.3\linewidth}{!}{
				\begin{tikzpicture}
				\begin{yquant}
                        
                        qubit {$q_1$} q;
                        qubit {$q_2$} q[+1];
                        qubit {$q_3$} q[+1];
                        qubit {$q_4$} q[+1];
                        qubit {$\vdots$} q[+1]; discard q[4];
                        qubit {$q_{n}$} q[+1];
                        qubit {$\ket0^{\otimes m}$} m;

				      h q[0-3];
				      h q[5];
                        box {$U^{2^0}$} (m[0]) | q[0];
                        box {$U^{2^1}$} (m[0]) | q[1];
                        text {$\dots$} q;
                        text {$\dots$} m;
                        
                        box {$U^{2^{n-1}}$} (m[0]) | q[5];
                        
                        box {$QFT^{\dagger}$} (q[0-5]);
				  \end{yquant}

				\end{tikzpicture}}};
				
				\node [right=of qc](est){
				\begin{tikzpicture}
				\node [align=center, draw](qubits) {\#Qubits \\ Runtime};
				\end{tikzpicture}};

				\node [above=of est](charac){
				\begin{tikzpicture}
				\node [align=center, label = {Assumed Hardware Characteristics}, draw]{Error Rates \\ Gate Execution Time \\ Error Budget \\ ...};
				\end{tikzpicture}};

				\node [right=of est](res){
				\begin{tikzpicture}

				\node [align=center](params) {Practicality \\ Assessment};
    				\node [right=of params, xshift=-10mm](checkmark) {\LARGE\color[RGB]{34,139,34}\checkmark};
				\end{tikzpicture}
    };
				
				\node[fit=(est)(res), draw, label=Estimation](est_box) {};

    \draw[-latex, line width=0.5mm] (p.east)  -- (designspace.west);
    \draw[-latex, line width=0.5mm] (designspace.east)  -- (qc.west);
    \draw[-latex, line width=0.5mm] (qc.east)  -- (est_box.west);
    \draw[-latex, line width=0.5mm] (est.east)  -- (res.west);
    \draw[-latex, line width=0.5mm] (charac.south)  -- (est.north);

    \draw[-latex, line width=0.5mm] (est.south)  |- ([yshift=-5mm] designspace.south) node[midway, below, xshift=-45mm] {Optimize design space parameters} --  (designspace.south);
				\end{tikzpicture}

}
	
	\vspace*{-4mm}
	\subfloat[Problem (taken from \cite{PhysRevResearch.3.033055}).\label{fig:wf2_1} ]{\hspace{.20\linewidth}}
	\hspace{0mm}
	\subfloat[Design Space.\label{fig:wf2_2} ]{\hspace{.20\linewidth}}
	\hspace{10mm}
	\subfloat[Quantum Program.\label{fig:wf2_3} ]{\hspace{.14\linewidth}}
	\hspace{6mm}
	\subfloat[Resource Estimation.\label{fig:wf2_4} ]{\hspace{.17\linewidth}}
	\hspace{0mm}
	\subfloat[Result.\label{fig:wf2_5} ]{\hspace{.15\linewidth}}
	
 \vspace{-1mm}
 \caption{Resource estimation-driven development
of quantum computing applications.}
\label{fig:proposed}
 \vspace{-7mm}
\end{figure*}

\subsection{Resource Estimation for Scalability Exploration}\label{sec:wf1}
\autoref{fig:proposed} illustrates how we integrate resource estimation into the workflow by substituting the \emph{Execution} step (indicated in \autoref{fig:wf1_4}/\autoref{fig:wf1_5}) by an \emph{Estimation} step (indicated in \autoref{fig:wf2_4}/\autoref{fig:wf2_5}).
More precisely, instead of returning the solution of the considered application instance, the complemented workflow returns an estimate of the required resources such as how many physical qubits would be needed for execution together with the estimated runtime.

\begin{example}
Consider again the problem from \autoref{ex:problem} and its respective quantum program derived in \autoref{ex:exec}.
Although this program represents a small problem instance that is easily executable on current NISQ computers, VQE generally does not scale well with respect to the size of the problem considered compared to other approaches, e.g., based on \emph{Quantum Phase Estimation}~(QPE)~\cite{liu2022prospects}.
Hence, larger problem instances such as the one depicted in \autoref{fig:wf2_1} would lead to different design space parameter choices (sketched in \autoref{fig:wf2_2})---e.g., \mbox{QPE-based} algorithms---and, thus, result in a different quantum program as shown in \autoref{fig:wf2_3}.
However, those approaches assume a (\mbox{close-to}) \mbox{error-free} execution---requiring error correction schemes to overcome the erroneous hardware and, thus, control the errors during computation in a way to stay within a provided error budget.
Using RE, the number of required physical qubits and the corresponding runtime can be determined as depicted in \autoref{fig:wf2_4} based on assumptions on the hardware characteristics of the qubits.
This ultimately leads to an assessment of whether a practical quantum advantage is likely for the provided problem as indicated in \autoref{fig:wf2_5}.
\end{example}

Following this approach gives an estimate of the required resources for \mbox{not-yet-executable} problem sizes. 
The resulting estimates of, e.g., required physical qubits and the runtime, might give an indication of the time horizon needed to solve those problem instances using actual quantum computers by comparing the required estimated resources with the roadmaps of device vendors.
This does not give an actual problem result, but a reference whether the considered problem is worth considering in the foreseeable future---an information highly relevant for various stakeholders from application developers in industry up to decision makers in funding agencies and politics to, e.g., define their national quantum strategy.

\subsection{Resource Estimation for Design Space Exploration}\label{sec:wf2}
Often, problems can be solved using different quantum computing approaches.  
For example, there exist various algorithms for simulating quantum chemistry, based on Trotterization~\cite{campbell2019simulation}, qubitization~\cite{Low2019hamiltonian}, or tensor hypercontraction~\cite{joonho2021thc}, to name a few.
On top of that, there are usually various possibilities to implement the quantum algorithm as a quantum program---and these choices have a significant influence on the required resources.
Therefore, a workflow that supports \mbox{end-users} in exploiting these design choices would certainly be helpful.

Using RE, the workflow described above can even be further complemented to provide this support for the design space exploration for a given problem by introducing a feedback loop as depicted in \autoref{fig:proposed} from \autoref{fig:wf2_4} to \autoref{fig:wf2_2}.
This loop allows \mbox{end-users} to determine the most promising set of design space parameters in a guided fashion. 
Finding such a set helps to assess the practicality of the execution of an algorithm instance on a future computer as sketched in~\autoref{fig:wf2_5}.

\begin{example} \label{ex:design_exploration}
Although the asymptotic computational complexity of quantum algorithms for simulating quantum chemistry can be compared, implementing them as quantum programs for a specific instance may introduce overhead that could alter the ranking of the algorithms and favor one over the other.
Moreover, while a particular algorithm may be favorable for one set of instances, other algorithms may be preferable for other sets. When considering various implementations for building blocks used in the quantum program, such as quantum arithmetic or table lookup, the search space expands even further.
\end{example}

This approach requires an effort to generate the initial quantum program that solves the problem considered based on the selected design space parameters before evaluating its scalability.
However, this is not always necessary and often it is sufficient to know how the selected quantum algorithm scales in terms of logical counts such as the required number of qubits and the number of gates with the problem size.
Using these numbers is already sufficient to run a resource estimate---simplifying the design exploration process even further.

Following this approach provides powerful means to explore and evaluate the design space how best to solve the considered problem already today without having to wait for the availability of sufficiently large quantum devices.
This early start of developing sophisticated quantum computing applications for \mbox{real-world} problem instances accelerates the race to catch up with classical solutions that have been developed and optimized for decades---a process that will take years, and, therefore, should be started as early as possible.

\subsection{Resource Estimation for Hardware Characteristics Exploration}\label{sec:wf3}
Finally, not only does the design space for how to encode an application into a quantum program offer a large degree of freedom, the assumed hardware characteristics used as input for the RE does as well.
So far, this freedom has been utilized by current RE tools by providing \mbox{pre-defined} configurations (such as the Azure Quantum Resource Estimator~\cite{beverland2022assessing} provides multiple configurations for both \mbox{gate-based} and Majorana qubits) of assumed hardware characteristics.
Additionally, these characteristics can be modified by \mbox{end-users}---and, hence, opens up the possibility of evaluating how further progress in hardware development might affect the resource estimates.

\begin{example} \label{ex:hw_exploration}
Similarly to how classical computers have evolved dramatically from early \mbox{small-scale} to current systems, quantum computing hardware is also expected to improve over time in terms of the number of qubits available, their error rates, decoherence times, and others. 
By employing RE, various hypotheses, such as an increase in gate execution speed or a decrease in gate execution error, can be examined and assessed.
\end{example}

Evaluating different hypotheses on how the quantum hardware may develop significantly aids to create a deeper understanding of the scalability of the application considered and its design space parameters.
Therefore, this evaluation should also be incorporated in the workflow of how quantum computing solutions are derived---leading to more informed and, thus, better design space parameters.

\begin{table}[t]
\caption{Resource estimates.}
\label{tab:res}
\vspace{-1mm}
\centering
\begin{tabular}{lrrrr}
    \toprule
    & Distance & Factories & Phys.~qubits & Runtime \\ \midrule
G $\mu$s, $10^{-3}$ & 31 &  12 & 5.62M & 156 years \\
G $\mu$s, $10^{-4}$ & 15 &  12 & 1.31M &  75 years \\
\textbf{G ns,} $\mathbf{10^{-3}}$    & \textbf{31} & \textbf{14} & \textbf{6.04M} &  \textbf{38 days} \\
G ns, $10^{-4}$     & 15 & 15 & 1.47M &  18 days \\
M ns, $10^{-4}$     & 17 & 13 & 4.20M &  16 days \\
\textbf{M ns,} $\mathbf{10^{-6}}$     &  \textbf{9} & \textbf{14} & \textbf{1.30M} &   \textbf{8 days} \\ \bottomrule
\end{tabular}
 \vspace{-4mm}
\end{table}

\section{Case Study}\label{sec:use_cases}

In this section, we demonstrate various RE scenarios in order to showcase the benefits of the workflow described above for a practical quantum chemistry application.\footnote{The source code for this experiment can be found at \url{https://github.com/cda-tum/mqt-problemsolver}.}
We evaluate the resources to calculate the ground state energy of a Hamiltonian to chemical accuracy of 1 mHartree using the qubitization quantum simulation algorithm~\cite{Low2019hamiltonian} on top of a \mbox{double-factorized} representation of the Hamiltonian~\cite{PhysRevResearch.3.033055}.
The Hamiltonian describes the 64 electron and 56 orbital active space of one of the stable intermediates in the ruthenium-catalyzed carbon fixation cycle~\cite{C4SC02087A} shown in~\autoref{fig:wf2_1}.

\subsection{Scalability Exploration}
As a first experiment, we estimate the resources for six examples of qubit parameters which represent various regimes of interest~\cite[Table II]{beverland2022assessing} as shown in \autoref{tab:res}.
The label contains the operation times regime (either $\mu$s or ns), and the limiting error rate of its Clifford operations, and is prefixed by whether the physical instruction sets are either \mbox{gate-based} (G) or have Majorana (M) instruction sets.
To ensure a chemical accuracy of 1 mHartree, we provide an error budget of 1\% to the resource estimator.

Next, we focus on the data point `M ns, $10^{-6}$' as a base estimate and evaluate space/time \mbox{trade-offs} by allowing a longer runtime of the algorithm. 
In the base estimate, 17.97\% of the total number of physical qubits is used to run 14 T factories in parallel that will produce the required 270 billion T states over the runtime of the algorithm. 
By slowing down the execution of the algorithm using logical idle operations, we require fewer T factories to run in parallel to produce the same number of required T states. 
We analyze how the runtime and the number of overall physical qubits is affected by limiting the number of T factories to $F = 1, \dots, 14$.  
The results are shown in \autoref{tab:tfac}.

\begin{table}[t]
\caption{Influence of T factories.}
\label{tab:tfac}
\vspace{-1mm}
\centering
\begin{tabular}{rrrrr}
    \toprule
    Factories & Distance & Fraction & Phys.~qubits & Runtime \\ \midrule
    14 & 9 & 17.97\% & 1.30M &   8 days \\
    13 & 9 & 16.91\% & 1.28M &   9 days \\
    12 & 9 & 15.81\% & 1.26M &  10 days \\
    11 & 9 & 14.69\% & 1.25M &  11 days \\
    10 & 9 & 13.53\% & 1.23M &  12 days \\
     9 & 9 & 12.35\% & 1.21M &  13 days \\
     8 & 9 & 11.13\% & 1.20M &  15 days \\
     7 & 9 &  9.87\% & 1.18M &  17 days \\
     6 & 9 &  8.58\% & 1.16M &  19 days \\
     5 & 9 &  7.26\% & 1.15M &  23 days \\
     \textbf{4} & \textbf{9} &  \textbf{5.89\%} & \textbf{1.13M} &  \textbf{29 days} \\
     3 & 9 &  4.49\% & 1.11M &  39 days \\
     2 & 9 &  3.04\% & 1.10M &  58 days \\
     1 & 9 &  1.54\% & 1.08M & 116 days \\ \bottomrule
\end{tabular}
\vspace{-4mm}
\end{table}

Assuming that one month is still a reasonable runtime, we find that with 4 factories we are below that threshold but are able to save about 170,000 physical qubits compared to the base estimate.

Although the current quantum computing application development usually relies on available quantum simulators and NISQ computers, the proposed methodologies allow \mbox{end-users} to get at least a resource estimate---and, by that, an assessment of its practicality---for their problem instances closer to \mbox{real-world} scenarios.

\subsection{Design Space Exploration}
Next, we make hypothetical assumptions on what impact optimizations to space (number of qubits) and time (number of operations) on the logical abstraction layer may have---potentially caused by different design space parameters.
For this purpose, we first extract the logical \mbox{pre-layout} counts from the previous RE results.
These are independent of the qubit parameters and space and time constraints, and are:
1,318 qubits, 67,474,931,068 Toffoli gates, 96 T gates, 11,987,084 rotation gates in depth of 11,986,482, as well as 63,472,407,520 single-qubit measurements. 
In addition to the baseline (using the data point `M ns, 10$^{-6}$') we consider three scenarios:
1) $\tfrac12$/2: use half the number of qubits, but all gate and depth counts are double, 
2) 2/$\tfrac12$: use twice the number of qubits, but all gate counts and depths are halved, 
3) $\tfrac34$/$\tfrac34$: save 25\% both in qubit counts, gate counts, and depth.
The respective resulting estimates are shown in \autoref{tab:desspace}.

\begin{table}[t]
\caption{Influence of different design space parameters.}
\label{tab:desspace}
\vspace{-1mm}
\centering
\begin{tabular}{lrrrr}
    \toprule
    Scenario              & Distance & Factories & Phys.~qubits & Runtime \\ \midrule
    baseline              & 9 & 14 &   1.30M &  8 days \\
    $\tfrac12$/2          & 9 & 14 & 773.06k & 17 days \\
    2/$\tfrac12$          & 9 & 14 &   2.34M &  4 days \\
    $\tfrac34$/$\tfrac34$ & 7 & 18 & 804.11k &  5 days \\ \bottomrule
\end{tabular}
 \vspace{-4mm}
\end{table}

As we can see, hypothetical savings do not fully propagate to the physical estimates, which emphasizes the importance of \mbox{full-stack} physical resource estimates. 
For the $\tfrac12$/2, the physical qubits did not decrease by 50\%, even though the runtime increased by more than twice.
On the other hand, for the 2/$\tfrac12$, the number of physical qubits did not increase by 2, even though the logical number of qubits is twice as high as the baseline.
Finally, in the optimistic scenario, both space and time savings are possible.
Since the savings allow for decreasing the code distance from 9 to 7, final savings of more
than 25\% to physical qubits and runtime are possible.

Even without the availability of sufficiently large quantum devices, different design space parameters can already be explored and their influences can be quantified and assessed.
Consequently, this allows \mbox{end-users} to keep improving their quantum computing applications such that optimized solutions are at hand once the corresponding quantum computing devices become available.

\subsection{Hardware Characteristics Exploration}
In the final experiment, we want to model how a potential change in error rates would affect the resulting resource estimates.
To this end, we start with the data point `G ns, $10^{-3}$' for a base estimate.
Here, the assumptions for the qubit parameters are $t_{\mathrm{meas}}^{(0)} = 100$ ns measurement time, and $t_{\mathrm{gate}}^{(0)} = 50$ ns gate operation times, as well $p^{(0)}_T = 10^{-3}$ measurement and gate error rates~\cite{beverland2022assessing}. 
We assume these parameters at some time point $t$ in the future, and model an improvement of $t_{\mathrm{meas}}^{(t)} = 0.9^t \cdot t_{\mathrm{meas}}^{(0)}$, $t_{\mathrm{gate}}^{(t)} = 0.9^t \cdot t_{\mathrm{gate}}^{(0)}$, and $p_T^{(t)} = 0.1^t \cdot p_T^{(0)}$ after $t$ equally spaced time steps.
These constants were arbitrarily chosen for the sake of the example and the resulting estimates are shown in \autoref{tab:hw_char}.
The experiments can be repeated using more accurate numbers from observing the rate of past improvements.

Using RE, the influence of such (rather simple) assumptions on the progress in quantum computing hardware can easily be determined for a given quantum program that realizes any application.
This provides end-users with powerful means to explore the design space for their problems---leading to more informed and, therefore, more scalable solutions.

\section{Discussion}\label{sec:discussion}
Although RE does not provide a solution to the problem considered, it can be used in a complementary fashion to the workflow of developing quantum computing applications.
RE is a powerful tool for exploring the scalability to \mbox{real-world} problem sizes under assumed hardware characteristics and error correction schemes---and, optionally, even considering hypotheses of how the hardware might improve.
By that, the most promising design space parameters can be determined although sufficiently large hardware is not yet available.
However, the approach is based on various assumptions, such as the assumed hardware characteristics and the methodology used to actually derive the resource estimates.
Due to the currently missing hardware on the required scale, it is impossible to validate those assumptions in an experimental evaluation to prove the methodology.

\begin{table}[t]
\caption{Influence of different hardware characteristics.}
\label{tab:hw_char}
\vspace{-1mm}
\centering
\begin{tabular}{lrrrr}
    \toprule
    $t$ & Distance & Factories & Phys.~qubits & Runtime \\ \midrule
0 & 31 & 14 &   6.04M & 38 days \\
1 & 15 & 15 &   1.47M & 17 days \\
2 & 11 & 13 & 733.28k & 11 days \\
3 &  7 & 14 & 295.96k &  6 days \\
4 &  7 & 10 & 278.52k &  6 days \\ \bottomrule
\end{tabular}
 \vspace{-4mm}
\end{table}

\section{Conclusions}\label{sec:conclusion}
Many of the current design flows for developing quantum computing applications focus on simulation and execution on near-term devices, which prohibits the exploration of \mbox{large-scale} quantum programs for quantum computing at scale.
This leads to the situation where quantum computing applications focus on \mbox{toy-size} problem instances without taking error correction into account.
In this paper, we describe how \emph{Resource Estimation} can readily complement such design flows.
Instead of execution a given quantum program, its required hardware resources are estimated.
Since this procedure does not rely on existing quantum simulators and computers, it is \emph{not} restricted to \mbox{toy-size} problem instances---resulting in three improvements:
(1)~\mbox{End-users} are enabled to consider \mbox{real-world} problem instances already today (also taking error correction schemes into account and determining the required hardware resources). 
These estimates can then be compared with the quantum hardware vendor's roadmaps to obtain a sense of the chosen approach's feasibility.
(2)~\mbox{End-users} can start exploring optimizations across the entire \emph{design space} of a quantum computing application to determine the most efficient parameters.
(3)~\mbox{End-users} can even incorporate various hypotheses of hardware development trends to derive more informed and, thus, better design space parameters.
Overall, the described workflow enables the development of quantum computing applications for \mbox{real-world} problem instances already today without having to wait for the availability of sufficiently large quantum devices.

\section*{Acknowledgments}
N.Q. and R.W. acknowledge funding from the European Research Council (ERC) under the European Union’s Horizon 2020 research and innovation program (DA QC, grant agreement No. 101001318 and MILLENION, grant agreement No. 101114305), the Munich Quantum Valley, which is supported by the Bavarian state government with funds from the Hightech Agenda Bayern Plus, and the BMWK on the basis of a decision by the German Bundestag through project QuaST, as well as the BMK, BMDW, the State of Upper Austria in the frame of the COMET program, and the QuantumReady project within Quantum Austria (managed by the FFG).

\vspace{5cm}

\clearpage
\printbibliography

\end{document}